\documentclass[times, 5p, sort&compress]{elsarticle}

\usepackage{array}
\usepackage{braket}
\usepackage{slashed}
\usepackage{graphicx}
\usepackage{subfig}
\usepackage{amsmath, amssymb, amsfonts}
\usepackage{mathtools}
\usepackage{mathrsfs}
  \DeclareMathAlphabet{\mathpzc}{OT1}{pzc}{m}{it}

\usepackage{textpos}
\newcommand{\D}[2]{\mathrm{d}^{#1}{#2}}

\makeatletter
  \def\ps@pprintTitle{%
       \let\@oddhead\@empty
       \let\@evenhead\@empty
       \def\@oddfoot{\footnotesize\itshape
         Preprint submitted to \@journal \hfill}%
     \let\@evenfoot\@oddfoot}
\makeatother

\journal{Physics Letters B}

\begin{document}

\begin{frontmatter}

\title{\vspace{-0.8cm}
{\bfseries Perturbative Non-Equilibrium Thermal Field Theory\\
    to all Orders in Gradient Expansion}}
\author{Peter Millington}
\ead{p.w.millington@shef.ac.uk}
\address{Consortium for Fundamental Physics,
  School of Mathematics and Statistics, \\ 
  University of Sheffield,
  Sheffield S3 7RH,
  United Kingdom}

\author{Apostolos Pilaftsis}
\ead{apostolos.pilaftsis@manchester.ac.uk}
\address{Consortium for Fundamental Physics,
  School of Physics and Astronomy, \\
  University of Manchester,
  Manchester M13 9PL,
  United Kingdom\vspace{-0.8cm}}

\begin{abstract}
We present  a new perturbative formulation  of non-equilibrium thermal
field  theory,   based  upon  non-homogeneous   free  propagators  and
time-dependent  vertices.  The  resulting  time-dependent diagrammatic
perturbation series  are free of pinch singularities  without the need
for  quasi-particle approximation or  effective resummation  of finite
widths.   After  arriving at  a  physically  meaningful definition  of
particle number  densities, we derive master  time evolution equations
for statistical distribution functions,  which are valid to all orders
in perturbation theory and to all orders in a gradient expansion.  For
a scalar model, we perform a perturbative loopwise truncation of these
evolution equations, whilst  still capturing fast transient behaviour,
which is found to  be dominated by energy-violating processes, leading
to the non-Markovian evolution of memory effects.
\end{abstract}


\end{frontmatter}

\begin{textblock}{4}(14.5,-12.8)
\begin{flushright}
\begin{footnotesize}
MAN/HEP/2013/06
\end{footnotesize}
\end{flushright}
\end{textblock}

\vspace{-3em}

\section{Introduction}

The  description   of  out-of-equilibrium  many-body  field-\linebreak
theoretic  systems  is  of  increasing relevance  in  theoretical  and
experimental physics  at the \emph{density  frontier}.  Examples range
from  the  early  Universe  to   the  deconfined  phase  of  QCD,  the
quark-gluon plasma, relevant at  heavy-ion colliders, such as RHIC and
the LHC  as well  as the internal  dynamics of  compact astro-physical
phenomena and condensed matter systems.

In  this Letter, we  present the  key concepts  of a  new perturbative
approach to non-equilibrium thermal quantum field theory, where master
time evolution equations for  macroscopic observables are derived from
first principles.  A comprehensive  exposition of this new formulation
is   provided    in   \cite{Millington:2012pf}.    In    contrast   to
semi-classical   approaches   based    on   the   Boltzmann   equation
\cite{Kolb:1979qa,  Carena:2002ss,  Giudice:2003jh,  Pilaftsis:2003gt,
  Buchmuller:2004nz,         Pilaftsis:2005rv,        Davidson:2008bu,
  Blanchet:2011xq},   this   new   approach  allows   the   systematic
incorporation of  finite-width and off-shell effects  without the need
for  effective  resummations.    Furthermore,  having  a  well-defined
underlying perturbation  theory that  is free of  pinch singularities,
these  time evolution  equations may  be truncated  in  a perturbative
loopwise  sense, whilst retaining  all orders  of the  time behaviour.
Several  studies appeared in  the literature~\cite{Danielewicz:1982kk,
  Lipavsky:1986zz,  Berera:1998gx,  Boyanovsky:1998pg, Niegawa:1999pn,
  Cassing:1999wx,   Dadic:1999bp,   Ivanov:1999tj,  Buchmuller:2000nd,
  Morawetz:1999bv,    Aarts:2001qa,   Blaizot:2001nr,   Juchem:2003bi,
  Prokopec:2003pj,           Prokopec:2004ic,           Berges:2004yj,
  Arrizabalaga:2005tf,          Berges:2005ai,         Lindner:2005kv,
  FillionGourdeau:2006hi,      DeSimone:2007rw,     Cirigliano:2007hb,
  Garbrecht:2008cb,  Garny:2009qn,  Cirigliano:2009yt,  Beneke:2010dz,
  Anisimov:2010dk,  Hamaguchi:2011jy,  Fidler:2011yq,  Gautier:2012vh,
  Drewes:2012qw}  proposing   quantum-corrected  transport  equations,
based  upon systems  of  Kadanoff--Baym equations~\cite{Kadanoff1989},
functional renormalization  group approaches \cite{Gasenzer:2010rq} or
expansion  of the  Liouville--von Neumann  equation \cite{Sigl:1992fn,
  Gagnon:2010kt}.  Whilst retaining all orders in perturbation theory,
the  existing approaches  often  rely on  the  truncation of  gradient
expansions  \cite{Winter:1986da, Bornath:1996zz} in  time derivatives,
quasi-particle approximations  or \emph{ad  hoc} ansaetze in  order to
obtain calculable  expressions or extract  meaningful observables.  In
this  new  perturbative  formalism, the  loopwise-truncated  evolution
equations  are   built  from  non-homogeneous   free  propagators  and
time-dependent  vertices.   This  diagrammatic approach  encodes  both
spatial  and temporal inhomogeneity  already from  tree-level, without
resorting to any such approximations.

\section{Canonical Quantization}
\label{sec:canquant}

We begin by highlighting the  details of the canonical quantization of
a  real  scalar  field   pertinent  to  a  perturbative  treatment  of
non-equilibrium thermal field theory.

The      \emph{time-independent}     Schr\"{o}dinger-picture     field
operator,\linebreak  denoted  by   a  subscript~$\mathrm{S}$,  may  be
written in the familiar plane-wave decomposition
\begin{equation}
  \Phi_{\mathrm{S}}(\mathbf{x};\tilde{t}_i)\ =\ \int\!\!
  \frac{\D{3}{\mathbf{p}}}{(2\pi)^3}\,\frac{1}{2E(\mathbf{p})}\;
  \Big(\,a_{\mathrm{S}}(\mathbf{p};\tilde{t}_i)e^{i\mathbf{p}\cdot\mathbf{x}}
  \: +\: a_{\mathrm{S}}^{\dag}(\mathbf{p};\tilde{t}_i)
  e^{-i\mathbf{p}\cdot\mathbf{x}}\,\Big)\; ,
\end{equation}
where   $E(\mathbf{p})\:   =\:    \sqrt{\mathbf{p}^2   +   M^2}$   and
$a_{\mathrm{S}}^{\dag}(\mathbf{p};          \tilde{t}_i)$          and
$a_{\mathrm{S}}(\mathbf{p};     \tilde{t}_i)$     are    the     usual
single-particle creation and  annihilation operators.  It is essential
to  emphasize  that  we  define the  Schr\"{o}dinger,  Heisenberg  and
interaction  (Dirac)   pictures  to   be  coincident  at   the  finite
\emph{micro}scopic boundary time~$\tilde{t}_i$, i.e.
\begin{equation}
  \Phi_{\mathrm{S}}(\mathbf{x};\tilde{t}_i)\ =\
  \Phi_{\mathrm{H}}(\tilde{t}_i,\mathbf{x};\tilde{t}_i)\ =\
  \Phi_{\mathrm{I}}(\tilde{t}_i,\mathbf{x};\tilde{t}_i)\; .
\end{equation}
It  is at  this picture-independent  boundary time  $\tilde{t}_i$ that
initial  conditions  must  be  specified.   The  dependence  upon  the
boundary  time $\tilde{t}_i$ is  separated from  other arguments  by a
semi-colon.

The       \emph{time-dependent}      interaction-picture      operator
$\Phi_{\mathrm{I}}(x;\tilde{t}_i)$   is  obtained   via   the  unitary
transformation
\begin{equation}
\Phi_{\mathrm{I}}(x;\tilde{t}_i)\:         =\:
e^{iH_{\mathrm{S}}^0(x_0\:              -\:              \tilde{t}_i)}
\Phi_{\mathrm{S}}(\mathbf{x};\tilde{t}_i)e^{-iH_{\mathrm{S}}^0(x_0\:
  -\: \tilde{t}_i)}\;,
\end{equation}
where $H_{\mathrm{S}}^0$  is the free  part of the Hamiltonian  in the
Schr\"{o}dinger picture. This yields
\begin{align}
  \label{eq:PhiI0}
  \Phi_{\mathrm{I}}(x;\tilde{t}_i)\ &=\ \!\int\!\!
  \frac{\D{3}{\mathbf{p}}}{(2\pi)^3}\,\frac{1}{2E(\mathbf{p})}\;
  \Big(\,a_{\mathrm{I}}(\mathbf{p},0;\tilde{t}_i)e^{-iE(\mathbf{p})x_0}
  e^{i\mathbf{p}\cdot \mathbf{x}}
  \nonumber\\& \qquad +\:
  a_{\mathrm{I}}^{\dag}(\mathbf{p},0;\tilde{t}_i)e^{iE(\mathbf{p})x_0}
  e^{-i\mathbf{p}\cdot\mathbf{x}}\,\Big)\; ,
\end{align}
where          $a_{\mathrm{I}}(\mathbf{p},x_0;\tilde{t}_i)\:         =
\:a_{\mathrm{I}}(\mathbf{p},0;\tilde{t}_i)\,e^{-iE(\mathbf{p})x_0}$
and   its   Hermitian    conjugate   are   the   \emph{time-dependent}
interaction-picture   annihilation  and  creation   operators.   These
operators satisfy the canonical commutation relation
\begin{align}
  \label{eq:momcomrel}
  &\big[\, a_{\mathrm{I}}(\mathbf{p},x_0;\tilde{t}_i\,),\
  a_{\mathrm{I}}^{\dag}(\mathbf{p}',x_0';\tilde{t}_i\,)\, \big]
  \nonumber\\&\qquad 
  =\ (2\pi)^3\,2E(\mathbf{p})\,\delta^{(3)}(\mathbf{p}\:-\:\mathbf{p}')\,
  e^{-iE(\mathbf{p})(x_0\:-\:x_0')}\; ,
\end{align}
with all other commutators vanishing.  Note the presence of an overall
phase  $e^{-iE(\mathbf{p})(x_0\:  -\: x_0')}$  in~(\ref{eq:momcomrel})
for $x_0 \neq x_0'$.

In quantum  statistical mechanics, we  are interested in  the Ensemble
Expectation Values  (EEVs) of operators at  a fixed \emph{micro}scopic
time of  observation $\tilde{t}_f$.  Such EEVs are  obtained by taking
the trace  with the density  operator $\rho(\tilde{t}_f;\tilde{t}_i)$,
i.e.
\begin{equation}
  \label{eq:EEV}
  \braket{\bullet}_{t}\ =\
  \mathcal{Z}^{-1}(t)\,\mathrm{Tr}\,
  \rho(\tilde{t}_f;\tilde{t}_i)\,\bullet\;,
\end{equation}
where                       $\mathcal{Z}(t)\:                      =\:
\mathrm{Tr}\,\rho(\tilde{t}_f;\tilde{t}_i)$ is the partition function,
which is time-dependent in the  presence of external sources.  We have
introduced            the            \emph{macro}scopic           time
$t\:=\:\tilde{t}_f\:-\:\tilde{t}_i$, which is the interval between the
\emph{micro}scopic boundary and observation times.

Consider  the  following observable,  which  is  the  EEV of  a
two-point product of field operators:
\begin{equation}
  \label{eq:obs}
  \mathcal{O}(\mathbf{x},\mathbf{y},\tilde{t}_f;\tilde{t}_i)\ =\ 
  \mathcal{Z}^{-1}(t)\,\mathrm{Tr}\,
  \rho(\tilde{t}_f;\tilde{t}_i)
  \Phi(\tilde{t}_f,\mathbf{x};\tilde{t}_i)
  \Phi(\tilde{t}_f,\mathbf{y};\tilde{t}_i)\;.
\end{equation}
As shown  in \cite{Millington:2012pf}, it is not  necessary to specify
the picture in which the operators of the RHS of (\ref{eq:obs}) are to
be  interpreted,  since all  operators  are  evaluated at  \emph{equal
  times}.   In  addition, the  observable  $\mathcal{O}$ is  invariant
under simultaneous  time translations of the  boundary and observation
times    and    depends   only    on    the   macroscopic    time~$t$:
$\mathcal{O}(\mathbf{x},\mathbf{y},\tilde{t}_f;\tilde{t}_i)\:    \equiv
\:\mathcal{O}(\mathbf{x},\mathbf{y},\tilde{t}_f-\tilde{t}_i;0)\:
\equiv     \:\mathcal{O}(\mathbf{x},\mathbf{y},t)$.     Notice    that
$\mathcal{O}$  depends  upon 7  independent  coordinates: the  spatial
coordinates  $\mathbf{x}$ and  $\mathbf{y}$ and  the  macroscopic time
$t$.

The    density   operator    $\rho(\tilde{t}_f;\tilde{t}_i)$    of   a
time-dependent and spatially  inhomogeneous background is non-diagonal
in the  Fock space and contains  an intractable incoherent  sum of all
possible    $n$    to    $m$    multi-particle    correlations,    see
\cite{Millington:2012pf}.   We may  account for  our ignorance  of the
exact form of this density operator by defining the bilinear EEVs
\begin{subequations}
\begin{align}
  \braket{a_{\mathrm{I}}(\mathbf{p},\tilde{t}_f;\tilde{t}_i)
    a_{\mathrm{I}}^{\dag}(\mathbf{p}',\tilde{t}_f;\tilde{t}_i)}_{t}\ &= \
  (2\pi)^3\,2E(\mathbf{p})\,\delta^{(3)}(\mathbf{p}-\mathbf{p}')\nonumber\\
  &\quad +\:2E^{\tfrac{1}{2}}(\mathbf{p})E^{\tfrac{1}{2}}(\mathbf{p}')
  f(\mathbf{p},\mathbf{p}',t)\;,\\
  \braket{a_{\mathrm{I}}^{\dag}(\mathbf{p}',\tilde{t}_f;\tilde{t}_i)
    a_{\mathrm{I}}(\mathbf{p},\tilde{t}_f;\tilde{t}_i)}_{t}\ &= \
  2E^{\tfrac{1}{2}}(\mathbf{p})E^{\tfrac{1}{2}}(\mathbf{p}')
  f(\mathbf{p},\mathbf{p}',t)\;,
\end{align}
\end{subequations}
consistent     with     the     canonical     commutation     relation
(\ref{eq:momcomrel}),                                             where
$f(\mathbf{p},\mathbf{p}',t)\:=\:f^*(\mathbf{p}',\mathbf{p},t)\,$.
The          \emph{statistical          distribution         function}
$f(\mathbf{p},\mathbf{p}',t)$  is  related   to  the  particle  number
density $n(\mathbf{q},\mathbf{X},t)$ via the Wigner transform
\begin{equation}
  \label{eq:nf}
  n(\mathbf{q},\mathbf{X},t)\ =\ 
  \!\int\!\!\frac{\D{3}{\mathbf{Q}}}{(2\pi)^3}\;
  e^{i\mathbf{Q}\cdot\mathbf{X}}\,
  f(\mathbf{q}+\mathbf{Q}/2,\mathbf{q}-\mathbf{Q}/2,t)\;,
\end{equation}
where   we  have   introduced   the  relative   and  central   momenta
$\mathbf{Q}\:=\:\mathbf{p}-\mathbf{p}'$                             and
$\mathbf{q}\:=\:(\mathbf{p}+\mathbf{p}')/2$, conjugate  to the central
and  relative  coordinates  $\mathbf{X}\:=\:(\mathbf{x}+\mathbf{y})/2$
and   $\mathbf{R}\:=\:\mathbf{x}-\mathbf{y}$,  respectively.   Observe
that  spatial homogeneity  is  broken by  the  explicit dependence  of
$f(\mathbf{p},\mathbf{p}',t)$  on the  two  three-momenta $\mathbf{p}$
and $\mathbf{p}'$.   In the  thermodynamic equilibrium limit,  we have
the    correspondence   $f(\mathbf{p},\mathbf{p}',t)\:   \rightarrow\:
f_{\mathrm{eq}}(\mathbf{p},\mathbf{p}')\:                           =\:
(2\pi)^3\,\delta^{(3)}(\mathbf{p}-\mathbf{p}')
f_{\mathrm{B}}\big(E(\mathbf{p})\big)$,                           where
$f_{\mathrm{B}}(x)\:=\:(e^{\beta  x}-1)^{-1}$  is  the  Bose--Einstein
distribution  function  and   $\beta$  is  the  inverse  thermodynamic
temperature.

\section{Schwinger--Keldysh CTP Formalism}
\label{sec:CTP}

We require a path-integral approach to generating EEVs for products of
field   operators.    Such   an    approach   is   provided   by   the
Schwinger--Keldysh      CTP      formalism     \cite{Schwinger:1960qe,
  Keldysh:1964ud}.

In order to obtain a generating functional of EEVs, we insert
unitary evolution operators to the left and right of the density
operator in the partition function
$\mathcal{Z}(t)\:=\:\mathrm{Tr}\,\rho(\tilde{t}_f;\tilde{t}_i)$, yielding
\begin{align}
  \label{eq:genfunc1}
  &\mathcal{Z}[\rho,J_{\pm},t]\nonumber\\ &\ \ \ = \: \mathrm{Tr}\,
    \Big[\bar{\mathrm{T}}
      e^{-i\!\int_{\Omega_t}\!\D{4}{x}\,J_-(x)\Phi_{\mathrm{H}}(x)}\Big]
    \,\rho_{\mathrm{H}}\big(\tilde{t}_f;\tilde{t}_i\big)\,
    \Big[\mathrm{T}
      e^{i\!\int_{\Omega_t}\!\D{4}{x}\,J_+(x)\Phi_{\mathrm{H}}(x)}\Big]\;,
\end{align}
in the Heisenberg picture,  where $\Omega_t$ is the temporally-bounded
spacetime  hypervolume  $[-t/2,\  t/2]\times\mathbb{R}^3$.  We  stress
that   (\ref{eq:genfunc1})   differs   fundamentally   from   existing
interpretations   of    the   CTP   formalism   \cite{Calzetta:1986ey,
  Calzetta:1986cq}.   Specifically,   the  Heisenberg-picture  density
operator    $\rho_{\mathrm{H}}(\tilde{t}_f;\tilde{t}_i)$,   which   is
explicitly  time-dependent in  the  presence of  the external  sources
$J_{\pm}$,   is   evaluated   at   the  \emph{time   of   observation}
$\tilde{t}_f$  and \emph{not}  the \emph{initial  time} $\tilde{t}_i$.
In our  approach, the  role of the  unitary evolution operators  is to
enable us to generate EEVs for products of field operators as given in
(\ref{eq:obs})  by  functional  differentiation  with respect  to  the
external sources.  The resulting  EEVs are evaluated at the \emph{time
  of observation}.

\begin{figure}
  \begin{center}
    \includegraphics[scale=0.85]{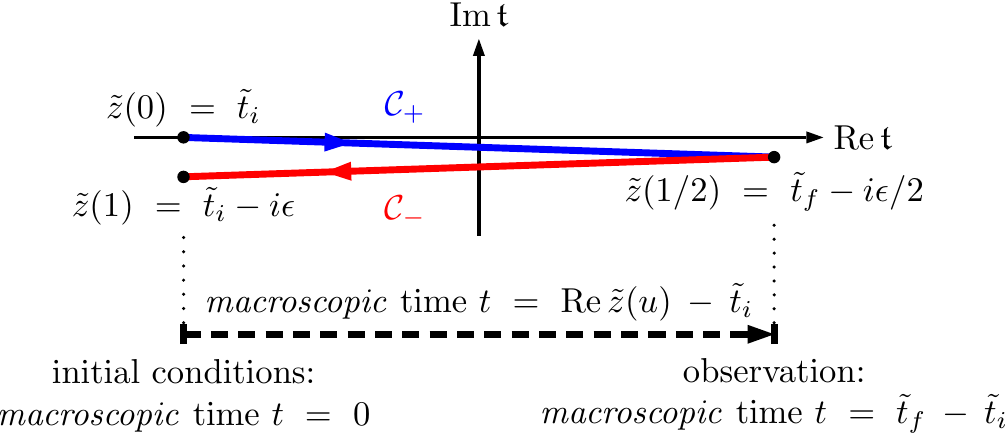}
    \vspace{-1em}
  \end{center}
  \caption{\label{fig:sk} The closed-time path,
    $\mathcal{C}\ =\ \mathcal{C}_+\:\cup\,\:\mathcal{C}_-$. The
    relationship between \emph{micro}scopic and \emph{macro}scopic
    times is indicated by a dashed black arrow.}
\end{figure}

We may  in\-terpret the evolution operators  in (\ref{eq:genfunc1}) as
de\-fining  a  closed  contour  ~$\mathcal{C}\: =\:  \mathcal{C}_+  \:
\cup\,\:     \mathcal{C}_-$      in     the     complex-time     plane
($\mathfrak{t}$-plane,  $\mathfrak{t}\:\in\:\mathbb{C}$), as  shown in
Figure~\ref{fig:sk}, which is the union of two anti-parallel branches:
$\mathcal{C}_+$,       running       from       $\tilde{t}_i$       to
$\tilde{t}_f\:-\:i\epsilon/2$;   and  $\mathcal{C}_-$,   running  from
$\tilde{t}_f\:-\:i\epsilon/2$  back to  $\tilde{t}_i\:-\:i\epsilon$. A
small imaginary part $\epsilon\:=\:0^+$  is added to separate the two,
essentially  co\-incident,  branches.  We  may  introduce an  explicit
parametrization       of       this       contour       $\tilde{z}(u)$
\cite{Millington:2012pf},  where  $u$  increases  monotonically  along
$\mathcal{C}$, which allows the definition of a path-ordering operator
$\mathrm{T}_{\mathcal{C}}$.  We emphasize that, in our formalism, this
contour evolves in time, with each branch having length $t$.

Following the notation of \cite{Calzetta:1986ey, Calzetta:1986cq}, we
denote fields confined to the positive and negative branches of the
CTP contour by $\Phi_{\pm}(x)\: \equiv\: \Phi(x^0 \in
\mathcal{C}_{\pm},\mathbf{x})$. We then define the doublets
\begin{subequations}
\begin{align}
  \Phi^a(x)\ &=\ \Big(\Phi_+(x)\,,\ \Phi_-(x)\Big)\;,\\
  \Phi_a(x)\ &=\ \eta_{ab}\Phi^b(x)\ =\ \Big(
  \Phi_+(x)\,,\ -\Phi_-(x)\Big)\; ,
\end{align}
\end{subequations}
where    the     CTP    indices    $a,\    b\:     =\:1,\    2$    and
$\eta_{ab}\:=\:\mathrm{diag}\,(1,\  -1)$ is an  $\mathbb{SO}\,(1,\ 1)$
`metric.' 

Inserting into (\ref{eq:genfunc1}) complete sets of eigenstates of the
Heis\-enberg field operator,  we derive a path-integral representation
of  the   CTP  generating  functional~\cite{Millington:2012pf},  which
depends on the path-ordered propagator
\begin{align}
  \label{eq:CTPprop}
  i\Delta^{ab}(x,y,\tilde{t}_f;\tilde{t}_i)\ &\equiv\ 
  \braket{\,\mathrm{T}_{\mathcal{C}}\,
    \big[\,\Phi^a(x;\tilde{t}_i)\Phi^b(y;\tilde{t}_i)\,\big]\,}_t
  \nonumber\\ &=\ 
  i\begin{bmatrix}
    \Delta_{\mathrm{F}}(x,y,\tilde{t}_f;\tilde{t}_i) &
    \Delta_{<}(x,y,\tilde{t}_f;\tilde{t}_i)
    \\ \Delta_{>}(x,y,\tilde{t}_f;\tilde{t}_i) &
    \Delta_{\mathrm{D}}(x,y,\tilde{t}_f;\tilde{t}_i)
  \end{bmatrix}\; .
\end{align}
For          $x^0,\           y^0\in{\cal          C}_+$,          the
path-ordering~$\mathrm{T}_{\mathcal{C}}$ is equivalent to the standard
time-ordering~$\mathrm{T}$  and  we  obtain the  time-ordered  Feynman
propagator~$i\Delta_{\mathrm{F}}(x,y,\tilde{t}_f;\tilde{t}_i)$.      On
the     other     hand,      for     $x^0,\     y^0\in{\cal     C}_-$,
$\mathrm{T}_{\mathcal{C}}$          is          equivalent          to
anti-time-ordering~$\bar{\mathrm{T}}$     and     we    obtain     the
anti-time-ordered                   Dyson                  propagator~
$i\Delta_{\mathrm{D}}(x,y,\tilde{t}_f;\tilde{t}_i)$.  For $x^0\in{\cal
  C}_+$ and $y^0\in{\cal C}_-$,  $x^0$ is always `earlier' than $y^0$,
yielding    the    absolutely-ordered   negative-frequency    Wightman
propagator\linebreak          $i\Delta_<(x,y,\tilde{t}_f;\tilde{t}_i)$.
Conversely, for  $y^0\in{\cal C}_+$ and $x^0\in{\cal  C}_-$, we obtain
the                     positive-frequency                    Wightman
propagator~$i\Delta_>(x,y,\tilde{t}_f;\tilde{t}_i)$.

By   means   of  a   Legendre   transform   of   the  CTP   generating
functional~\cite{Millington:2012pf},   we    derive   the   respective
Cornwall--Jackiw--Tomboulis  effective  action \cite{Cornwall:1974vz},
from which the CTP \linebreak Schwinger--Dyson equation
\begin{equation}
\label{eq:SD}
  \Delta_{ab}^{-1}(x,y,\tilde{t}_f;\tilde{t}_i)\ =\
  \Delta_{ab}^{0,-1}(x,y)\:+\:\Pi_{ab}(x,y,\tilde{t}_f;\tilde{t}_i)\;
\end{equation}
is obtained, where $\Delta_{ab}^{-1}(x,y,\tilde{t}_f;\tilde{t}_i)$ and
$\Delta_{ab}^{0,\,-1}(x,y)$  are  the re-\linebreak summed  and  free inverse  CTP
propagators, respectively, and $\Pi_{ab}(x,y,\tilde{t}_f;\tilde{t}_i)$
is the CTP self-energy, analogous in form to (\ref{eq:CTPprop}).

\section{Master Time Evolution Equations for \\
  Particle Number Densities}
\label{sec:evo}  

In  order   to  count   both  on-shell  and   off-shell  contributions
systematically, we `measure' the number of charges, rather than quanta
of energy. This avoids any need to identify `single-particle' energies
by means of  a quasi-particle approximation. We begin  by relating the
Noether charge
\begin{equation}
  \label{eq:Q}
  \mathcal{Q}(x_0;\tilde{t}_i)\ =\ -\,i\!\int\!\D{3}{\mathbf{x}}\;
  \Big(\,\pi_{\mathrm{H}}(x;\tilde{t}_i)\,\Phi_{\mathrm{H}}(x;\tilde{t}_i)
  \:-\:\mathrm{H.\ c.}\,\Big)
\end{equation}
to a charge density operator
$\mathcal{Q}(\mathbf{q},\mathbf{X},X_0;\tilde{t}_i)$ via
\begin{equation}
  \label{eq:Qdef}
  \mathcal{Q}(X_0;\tilde{t}_i)\ =\ \!\int\D{3}{\mathbf{X}}\int\!\!
  \frac{\D{3}{\mathbf{q}}}{(2\pi)^3}\;
  \mathcal{Q}(\mathbf{q},\mathbf{X},X_0;\tilde{t}_i)\;,
\end{equation}
where  $\pi_{\mathrm{H}}(x;\tilde{t}_i)$  is  the  conjugate  momentum
operator   to  $\Phi_{\mathrm{H}}(x;\tilde{t}_i)$.    By   taking  the
equal-time EEV of $\mathcal{Q}(\mathbf{q},\mathbf{X},X_0;\tilde{t}_i)$
and   extracting  the   positive-   and  negative-frequency   particle
components,  we arrive  at the  following definition  of  the particle
number     density    in     terms     of    off-shell     propagators
\cite{Millington:2012pf}:
\begin{align}
  n(\mathbf{q},\mathbf{X},t)\ &=\ \lim_{X_0\:\to\: t}\,2\!\int\!
  \frac{\D{}{q_0}}{2\pi}\!\int\!\!\frac{\D{4}{Q}}{(2\pi)^4}\;
  e^{-iQ\cdot X}\,\nonumber\\&\qquad \times\:\theta(q_0)q_0
  i\Delta_<(q+\tfrac{Q}{2},q-\tfrac{Q}{2},t;0)\;,
\end{align}
using the translational invariance of the CTP contour.

By   partially  inverting   the  CTP   Schwinger--Dyson   equation  in
(\ref{eq:SD}), we derive the  following master time evolution equation
for   the   statistical    distribution   function   $f(\mathbf{q}   +
\tfrac{\mathbf{Q}}{2},\mathbf{q}-\tfrac{\mathbf{Q}}{2},t)$
\cite{Millington:2012pf}:
\begin{align}
  \label{eq:evo}
  &\partial_{t}
  f(\mathbf{q}+\tfrac{\mathbf{Q}}{2},\mathbf{q}-\tfrac{\mathbf{Q}}{2},t)
  \nonumber\\&\qquad -\:
  2\!\iint\!\frac{\D{}{q_0}}{2\pi}\,\frac{\D{}{Q_0}}{2\pi}\;e^{-iQ_0t}
  \,\mathbf{q}\cdot\mathbf{Q}\,\theta(q_0)
  \Delta_<(q+\tfrac{Q}{2},q-\tfrac{Q}{2},t;0)\nonumber\\&
  \qquad +\: \!\iint\!\frac{\D{}{q_0}}{2\pi}\,
  \frac{\D{}{Q_0}}{2\pi}\;e^{-iQ_0t}\,\theta(q_0)
  \Big(\,\mathscr{F}(q+\tfrac{Q}{2},q-\tfrac{Q}{2},t;0)
  \nonumber\\& \qquad \qquad+\:
  \mathscr{F}^{*}(q-\tfrac{Q}{2},q+\tfrac{Q}{2},t;0)\,\Big)\nonumber\\&
  \qquad =\ \!\iint\!\frac{\D{}{q_0}}{2\pi}\,
  \frac{\D{}{Q_0}}{2\pi}\;e^{-iQ_0t}\,\theta(q_0)
  \Big(\,\mathscr{C}(q+\tfrac{Q}{2},q-\tfrac{Q}{2},t;0)
  \nonumber\\&\qquad \qquad +\:
  \mathscr{C}^{*}(q-\tfrac{Q}{2},q+\tfrac{Q}{2},t;0)\,\Big)\;,
\end{align}
where we have introduced
\begin{subequations}
\begin{align}
  \label{eq:fdef}
  &\mathscr{F}(q+\tfrac{Q}{2},q-\tfrac{Q}{2},t;0)\nonumber\\ &\qquad \equiv\ 
  -\!\int\!\!\frac{\D{4}{k}}{(2\pi)^4}\;
  i\Pi_{\mathcal{P}}(q+\tfrac{Q}{2},k,t;0)\:
  i\Delta_<(k,q-\tfrac{Q}{2},t;0)\;,\\
  \label{eq:cdef}
  &\mathscr{C}(q+\tfrac{Q}{2},q-\tfrac{Q}{2},t;0)\nonumber\\ &\qquad \equiv\
  \frac{1}{2}\!\int\!\!\frac{\D{4}{k}}{(2\pi)^4}\;
  \Big[\,i\Pi_>(q+\tfrac{Q}{2},k,t;0)\:
  i\Delta_<(k,q-\tfrac{Q}{2},t;0)
  \nonumber\\&\qquad
  -\: i\Pi_<(q+\tfrac{Q}{2},k,t;0)\,
  \Big(\,i\Delta_>(k,q-\tfrac{Q}{2},t;0)
  \nonumber\\&\qquad
  \qquad-\:2i\Delta_{\mathcal{P}}(k,q-\tfrac{Q}{2},t;0)\,\Big)\Big]\;. 
\end{align}
\end{subequations}
It  is   important  to   emphasize  that  (\ref{eq:evo})   provides  a
self-consistent  time evolution  equation for~$f$  valid  \emph{to all
  orders} in perturbation theory  and to \emph{all orders} in gradient
expansion. The  terms on the  LHS of (\ref{eq:evo}) may  be associated
with     the     total    derivative     in     the    phase     space
$(\mathbf{X},\ \mathbf{p})$, which  appears in the classical Boltzmann
transport equation \cite{KolbTurner}.  The expression $\mathscr{F}$ in
(\ref{eq:fdef}) is  the \emph{force} term, generated  by the potential
due to the  dispersive part of the self-energy,  and the $\mathscr{C}$
in (\ref{eq:cdef}) are the \emph{collision} terms.

\section{Non-Homogeneous Diagrammatics}\label{sec:nonhom}

Let us  consider a  simple scalar theory,  with one heavy  real scalar
field   $\Phi$  and   one  light   pair  of   complex   scalar  fields
$(\chi^{\dag}$, $\chi)$, described by the Lagrangian
\begin{equation}
  \label{eq:model}
  \mathcal{L}\ =\ \tfrac{1}{2}\partial_{\mu}\Phi\partial^{\mu}
  \Phi\:-\:\tfrac{1}{2}M^2\Phi^2\:+\:\partial_{\mu}
  \chi^{\dag}\partial^{\mu}\chi\:-\:m^2\chi^{\dag}\chi\:-\: 
  g\Phi\chi^{\dag}\chi\:-\:\cdots\;,
\end{equation}
where  the  ellipsis contains  omitted  self-interactions. This  model
yields the following set of modified Feynman rules:
\begin{table*}
  \begin{center}
    \begin{tabular}{| m{3.5cm} | m{12.75cm} |}
      \hline ~ & ~
      \\
      {\bfseries Propagator} & {\bfseries Double-Momentum Representation}
      \\
      \hline
      Feynman (Dyson) &
      $\begin{array}{l}
        \vspace{-0.8em} \\
        i\Delta^0_{\mathrm{F}(\mathrm{D})}(p,p',\tilde{t}_f;\tilde{t}_i)\ =\
        \displaystyle \frac{(-)i}{p^2-M^2+(-)i\epsilon}
        (2\pi)^4\delta^{(4)}(p-p')\nonumber\\\qquad \qquad \qquad \qquad \qquad 
        +\:2\pi|2p_0|^{1/2}\delta(p^2-M^2)\tilde{f}(p,p',t)
        e^{i(p_0-p_0')\tilde{t}_f}2\pi|2p_0'|^{1/2}\delta(p'^2-M^2)
      \vspace{0.3em}\end{array}$ \\\hline
      $+$($-$)ve-freq. Wightman &
      $\begin{array}{l}
        \vspace{-0.7em} \\
        i\Delta_{>(<)}^0(p,p',\tilde{t}_f;\tilde{t}_i)\ =\
        2\pi\theta(+(-)p_0)\delta(p^2-M^2)(2\pi)^4\delta^{(4)}(p-p')\\
        \qquad \qquad \qquad \qquad \qquad +\:2\pi|2p_0|^{1/2}\delta(p^2-M^2)
        \tilde{f}(p,p',t)e^{i(p_0-p_0')\tilde{t}_f}2\pi|2p_0'|^{1/2}\delta(p'^2-M^2)
      \vspace{0.3em}\end{array}$\\\hline 
      Retarded (Advanced) &
      $\begin{array}{l}
        \vspace{-0.7em} \\
        i\Delta_{\mathrm{R}(\mathrm{A})}^0(p,p')\ =\
        \displaystyle \frac{i}{(p_0+(-)i\epsilon)^2-\mathbf{p}^2-M^2}
        (2\pi)^4\delta^{(4)}(p-p')
      \vspace{0.3em}\end{array}$ \\\hline
      Pauli--Jordan &
      $\begin{array}{l}
        \vspace{-0.7em} \\
        i\Delta^0(p,p')\ =\ 2\pi\varepsilon(p_0)
        \delta(p^2-M^2)(2\pi)^4\delta^{(4)}(p-p')
      \vspace{0.3em}\end{array}$\\\hline
      Hadamard &
      $\begin{array}{l}
        \vspace{-0.7em} \\
        i\Delta_1^0(p,p',\tilde{t}_f;\tilde{t}_i)\ =\
        2\pi\delta(p^2-M^2)(2\pi)^4\delta^{(4)}(p-p')\\
        \qquad \qquad \qquad \qquad \qquad+\:2\pi|2p_0|^{1/2}\delta(p^2-M^2)2\tilde{f}(p,p',t)
        e^{i(p_0-p_0')\tilde{t}_f}2\pi|2p_0'|^{1/2}\delta(p'^2-M^2)
      \vspace{0.3em}\end{array}$\\\hline
      Principal-part &
      $\begin{array}{l}
        \vspace{-0.7em} \\
        \displaystyle i\Delta^0_{\mathcal{P}}(p,p')\ =\
        \mathcal{P}\frac{i}{p^2-M^2}(2\pi)^4\delta^{(4)}(p-p')
        \vspace{0.4em} 
      \end{array}$\\\hline
      \end{tabular}
  \end{center}
  \vspace{-1em}
  \caption{The non-homogeneous free scalar propagators, where
    $\tilde{f}(p,p',t)\: =\:
    \theta(p_0)\theta(p_0')f(\mathbf{p},\mathbf{p}',t)\:
    +\:\theta(-p_0)\theta(-p_0')f^*(-\mathbf{p},-\mathbf{p}',t)$,
    $\theta(p_0)$ is the unit step function and $\varepsilon(p_0)$ is
    the signum function.}
  \label{tab:nonhom}
\end{table*}

\begin{itemize}

\item sum over all topologically distinct diagrams at a given order in
  perturbation theory.

\item assign to each $\Phi$-propagator line a factor of
\begin{equation*}
\parbox[][17.5pt][t]{80pt}{\vspace{-0.75em}\centering
  \includegraphics[scale=0.75]{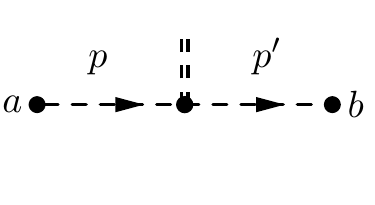}}\ =\
  i\Delta^{0,\,ab}_{\Phi}(p,p',\tilde{t}_f;\tilde{t}_i)\;.
\end{equation*}
The set of non-homogeneous free propagators is listed in Table
\ref{tab:nonhom}.

\item assign to each $\chi$-propagator line a factor of
\begin{equation*}
\parbox[][17.5pt][t]{80pt}{\vspace{-0.75em}\centering
  \includegraphics[scale=0.75]{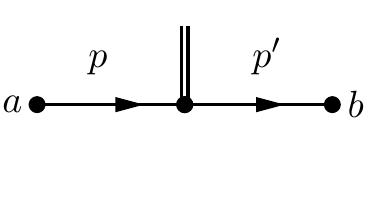}}\ =\
  i\Delta^{0,\,ab}_{\chi}(p,p',\tilde{t}_f;\tilde{t}_i)\;.
\end{equation*}
The  double  lines  occurring  in  the  CTP  propagators  reflect  the
violation of three-momentum due to the non-homogeneous
statistical distribution function $f(\mathbf{p},\mathbf{p}',t)$.

\item assign to each three-point vertex a factor of
\begin{equation*}
  \parbox[][70pt][c]{80pt}{\centering
  \includegraphics[scale=0.7]{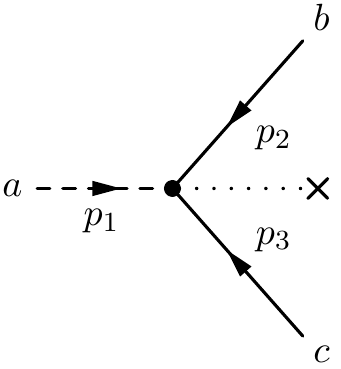}}\ =\ -ig\eta_{abc}\,(2\pi)^4
  \delta^{(4)}_t\big(\textstyle \sum_{i=1}^3p_i\big)\;,
\end{equation*}
where $\eta_{abc\cdots}\:=\:1$, $a\:=\:b\:=\:\cdots\:=\:1$;
$\eta_{abc\cdots}\:=\:-1$, $a\:=\:b\:=\:\cdots\:=\:2$ and
$\eta_{abc\cdots}\:=\:0$ otherwise. Due to the finite upper and lower
bounds on the interaction-dependent time integrals, the
energy-momentum delta function is replaced by
\begin{equation}
  \delta^{(4)}_{t}\big(\textstyle\sum_{i=1}^3p_i\big)\ \equiv \ 
  \delta_{t}\big(\textstyle\sum_{i=1}^3p_{0,\,i}\big)
  \delta^{(3)}(\mathbf{p}-\mathbf{p}')
\end{equation}
in  which  energy  conservation  is  systematically  violated  by  the
analytic weight function
\begin{equation}
  \label{eq:deltat}
  \delta_t\big({\textstyle \sum_{i=1}^3p_{0,\,i}}\big)\ \equiv \
  \frac{t}{2\pi}\,\mathrm{sinc}\,
  \big[\big({\textstyle \sum_{i=1}^3p_{0,\,i}}\big)t/2\big]\;.
\end{equation}
This violation of energy conservation, shown diagrammatically by the
dotted line terminated in a cross, results from the uncertainty
principle, since the observation of the system is made over a finite
time interval. We ignore additional statistical contributions to
vertices that result from a possible non-Gaussian density operator
(for a discussion, see \cite{Millington:2012pf}).

\item  associate  with  each external vertex a phase
\begin{equation*}
  e^{ip_0\tilde{t}_f}\;,
\end{equation*}
where $p_0$ is  the energy flowing \emph{into} the  vertex. This phase
results  from the  proper consideration  of the  Wick  contraction and
field-particle duality relations.

\item contract all internal CTP indices.

\item integrate with the measure
\begin{equation*}
  \int\!\!\frac{\D{4}{p}}{(2\pi)^4}\;
\end{equation*}
over  the four-momentum associated  with each contracted pair  of CTP
indices.

\item consider the combinatorial symmetry factors, where appropriate.
\end{itemize}
These non-homogeneous Feynman rules encode the absolute spacetime
dependence of the system starting from \emph{tree level}.

\section{Absence of Pinch Singularities}

The perturbation  series built from the  non-homogeneous Feynman rules
in  Section  \ref{sec:nonhom}  are  free of  the  pinch  singularities
previously   thought  to   spoil  such   perturbative   treatments  of
non-equilibrium    field   theory,    see   e.g.~\cite{Altherr:1994fx,
  Altherr:1994jc,Greiner:1998ri,Berges:2004yj}.   In  our formulation,
this absence of pinch singularities is ensured by two factors: (i) the
violation  of  energy  conservation   at  early  times  and  (ii)  the
statistical  distribution  functions   in  free  CTP  propagators  are
evaluated at  the \emph{time of  observation}.  The latter (ii)  is in
contrast to existing approaches in which {\em free} propagators do not
evolve and depend only on the \emph{initial} distributions.

Consider the following one-loop insertion to the propagator: 
\begin{align}
  &i\Delta^{(1),\, ab}(p,p',\tilde{t}_f;\tilde{t}_i)\ =\
  i\Delta^{0,\, ac}(p,p',\tilde{t}_f;\tilde{t}_i)\nonumber\\
  &\ +i\Delta^{0,\,ac}(p,q,\tilde{t}_f,\tilde{t}_i)
  i\Pi^{(1)}_{cd}(q,q',\tilde{t}_f;\tilde{t}_i)
  i\Delta^{0,\,db}(q',p',\tilde{t}_f,\tilde{t}_i)\;.
\end{align}
Potential pinch singularities arise from terms like
\begin{equation}
  \delta(p^2-M^2)\delta(p_0-p_0')\delta(p'^2-M^2)\;.
\end{equation}
However, at early times, energy is not conserved through the loop
insertion. As a result, these terms are analytic, becoming
\begin{equation}
  \delta(p^2-M^2)\delta_t(p_0-p_0')\delta(p'^2-M^2)\;,
\end{equation}
where $\delta_t(p_0-p_0')$ is given in (\ref{eq:deltat}). At late
times, $t\:\to\:\infty$,
\begin{equation}
  \lim_{t\to\infty}\,\delta_t(p_0-p_0')\ =\ \delta(p_0-p_0')
\end{equation}
and energy conservation  is restored.  However, in the  same limit the
system  must   have  thermalized.   In  this   case,  the  statistical
distribution  functions  appearing in  free  propagators  will be  the
equilibrium distributions  for which pinch singularities  are known to
cancel  by  virtue   of  the  Kubo--Martin--Schwinger  (KMS)  relation
\cite{LeBellac2000}.  At intermediate  times, pinch singularities grow
like a power law in $t$,  which will always occur more slowly than the
exponential approach to equilibrium.  Thus, the perturbation series is
free of pinch singularities for all times~\cite{Millington:2012pf}.

Given the systematic diagrammatics  of this approach, we may therefore
truncate the  master time evolution  equations in (\ref{eq:evo})  in a
perturbative   loopwise  sense.    If  the   statistical  distribution
functions are tempered for all times, any ultra-violet divergences may
be  renormalized by the  usual zero-temperature  counter-terms, whilst
infra-red divergences may be regularized by the partial resummation of
thermal masses, see \cite{Millington:2012pf}.

\section{Time-Dependent One-Loop Width}

\begin{figure}
  \begin{center}
    \vspace{-0.75em}
    \includegraphics[scale=0.75]{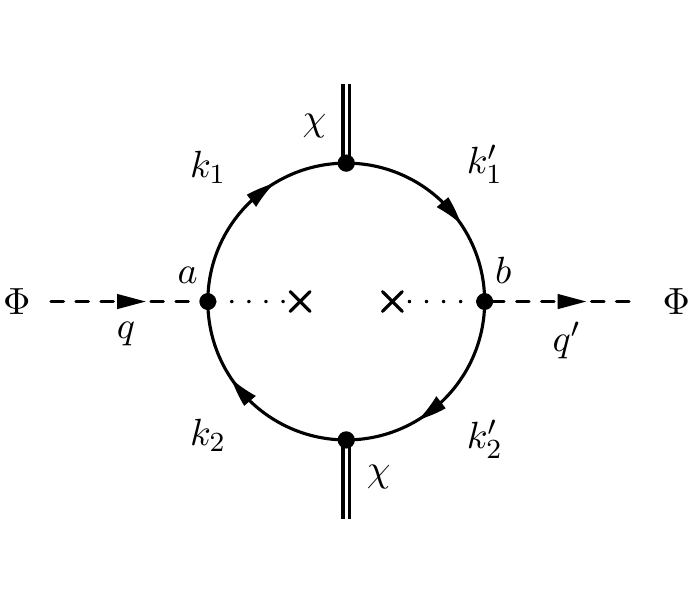}
    \vspace{-2.5em}
  \end{center}
  \caption{The one-loop $\Phi$ self-energy
    $i\Pi^{(1)}_{\Phi,\,ab}(q,q',\tilde{t}_f;\tilde{t}_i)$.}
  \label{fig:selfs}
\end{figure}

To illustrate the distinctive  features of our perturbative formalism,
let   us    consider   two   isolated    but   coincident   subsystems
$\mathscr{S}_{\Phi}$  and  $\mathscr{S}_{\chi}$,  both  separately  in
thermodynamic    equilibrium    and    at   the    same    temperature
$T=10~\mathrm{GeV}$ with the interactions switched off.  The subsystem
$\mathscr{S}_{\Phi}$  contains   only  the  field   $\Phi$  with  mass
$M=1~\mathrm{GeV}$ and $\mathscr{S}_{\chi}$, only the $\chi$ fields of
mass $m=0.01~\mathrm{GeV}$.  At $t=0$, we turn on the interactions and
allow           the           system          $\mathscr{S}           =
\mathscr{S}_{\Phi}\cup\mathscr{S}_{\chi}$ to re-thermalize.

\begin{figure}
  \begin{center}
    \subfloat[$1\rightarrow 2$ decay]{\label{fig:kinsa}
      {\centering \qquad \includegraphics[scale = 0.75]{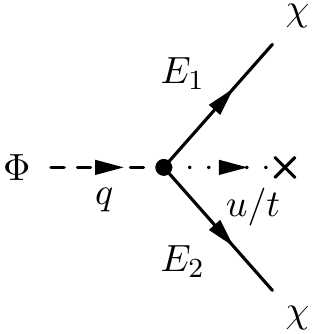}\qquad }}
    \subfloat[$3\rightarrow 0$ total annihilation]{\label{fig:kinsc}
      {\centering  \qquad \includegraphics[scale = 0.75]{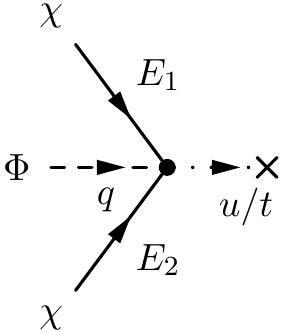}\qquad }}
      \\
    \subfloat[$2\rightarrow 1$ Landau damping]{\label{fig:kinsb}
      {\centering \includegraphics[scale = 0.75]{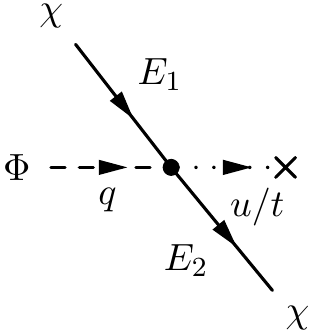} \
        \includegraphics[scale = 0.75]{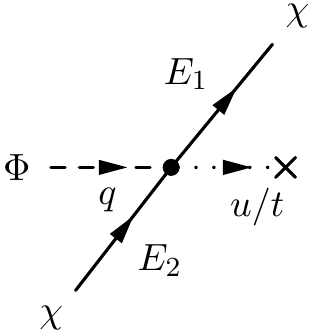}}}\\
    \vspace{-0.5em}
  \end{center}
  \caption{\label{fig:kins} The four evanescent processes contributing to the
    one-loop time-dependent $\Phi$ width.}
\end{figure}

The    one-loop   non-local   $\Phi$    self-energy   is    shown   in
Figure~\ref{fig:selfs}.   Neglecting  back-reaction  on the  subsystem
$\mathscr{S}_{\chi}$, the one-loop time-dependent $\Phi$ width is then
given by the following integral:
\begin{align}
  \label{eq:phiwidth}
  &\Gamma^{(1)}_{\Phi}(q,t)\         =\        \frac{g^2t}{64\pi^3M}\!
  \sum_{\alpha_{1},\,\alpha_{2}\:=\:\pm     1}\int\!\D{3}{\mathbf{k}}\;
  \frac{\alpha_1\alpha_2}{E_1E_2}\,\nonumber\\&\qquad          \times\:
  \mathrm{sinc}\big[\big(q_0-\alpha_1E_1-\alpha_2E_2\big)\,t\big]
  \,\big(1+f_{\mathrm{B}}(\alpha_1E_1)+
  f_{\mathrm{B}}(\alpha_2E_2)\big)\;,
\end{align}  
where              $E_1\:\equiv\:              E_{\chi}(\mathbf{k})\:=
\:\sqrt{\mathbf{k}^2+m^2}$                                          and
$E_2\:\equiv\:E_{\chi}(\mathbf{q}-\mathbf{k})$.    The   violation  of
energy conservation, due  to the sinc function in~(\ref{eq:phiwidth}),
leads       to       otherwise-forbidden      contributions       from
$\alpha_1,\     \alpha_2\:=\:-1$      (total     annihilation)     and
$\alpha_1\:=\:-\alpha_2$   (Landau   damping).    In   addition,   the
kinematically-allowed phase  space for $1  \to 2$ decays  is expanded.
These evanescent  processes are shown  in Figure~\ref{fig:kins}, where
we have defined the evanescent action
\begin{equation}
  \label{eq:u}
  u\ \equiv\ (q_0-\alpha_1E_1-\alpha_2 E_2)\,t\;,
\end{equation}
quantifying    the   degree    of    energy   non-conservation.    For
$t\:\to\:\infty$, we recover the known equilibrium result, since
\begin{equation}
  \label{eq:tinf}
  \lim_{t\:\to\:\infty}\,\frac{t}{\pi}\,\mathrm{sinc}\big[
  \big(q_0-\alpha_1E_1-\alpha_2
  E_2\big)t\,\big]\ =\
  \delta\big(q_0-\alpha_1E_1-
  \alpha_2E_2\big)\;.
\end{equation}

\begin{figure*}
  \begin{center}
    \includegraphics[scale=0.85]{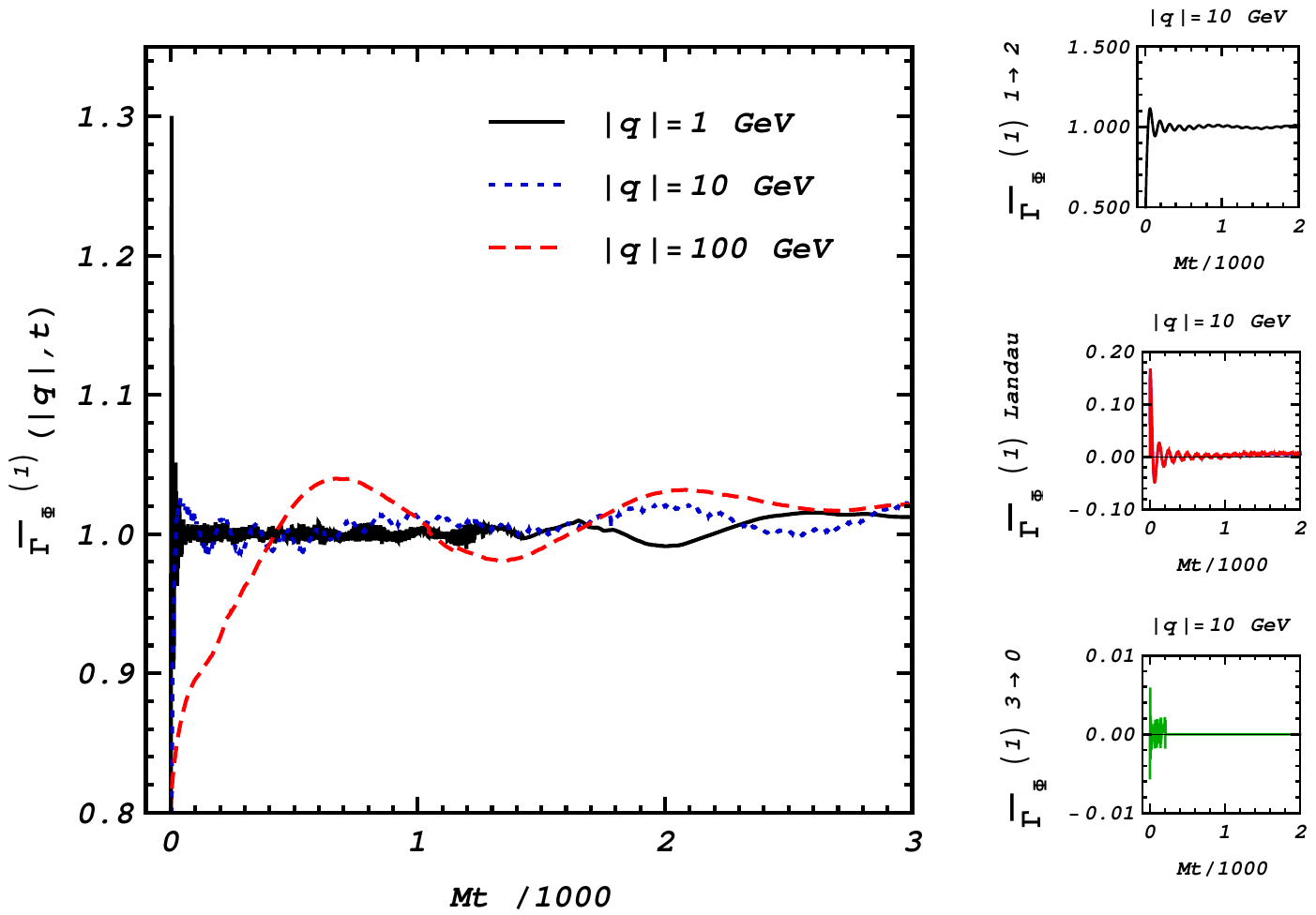}
    \vspace{-1em}
  \end{center}
  \caption{\emph{Left}: the ratio $\bar{\Gamma}_{\Phi}^{(1)}$ versus
    $Mt$ for on-shell decays with $|\mathbf{q}|\:=\:1\ \mathrm{GeV}$
    (solid black), $10\ \mathrm{GeV}$ (blue dotted) and
    $100\ \mathrm{GeV}$ (red dashed). \emph{Right}: separate
    contributions to $\bar{\Gamma}_{\Phi}^{(1)}$ for
    $|\mathbf{q}|\:=\:10\ \mathrm{GeV}$. Landau--damping
    contributions are equal up to numerical errors. }
  \label{fig:twidth}
\end{figure*}

In Figure~\ref{fig:twidth}, we plot the ratio
\begin{equation}
  \bar{\Gamma}_{\Phi}^{(1)}(|\mathbf{q}|,t)\ =\ 
  \frac{\Gamma_{\Phi}^{(1)}(|\mathbf{q}|,t)}{
  \Gamma_{\Phi}^{(1)}(|\mathbf{q}|,t\to\infty)}
\end{equation}
of  the   time-dependent  one-loop  $\Phi$  width   to  its  late-time
equilibrium value as  a function of $Mt$ for $q^2\:=\:M^2$.
In addition, we plot the separate contributions of the processes shown
in Figure~\ref{fig:kins}.

\section{Non-Markovian Oscillations}

In Figure \ref{fig:twidth}, we observe that the oscillations in the
$\Phi$ width have time-dependent frequencies.  This non-Markovian
beha\-viour is inherent to truly out-of-equilibrium quantum systems,
exhibiting so-called \emph{memory effects}.  Moreover, due to the
\linebreak Lorentz boost of ultra-violet modes relative to the rest
frame of the heat bath, these memory effects persist for timescales
much longer than the $1/M$ that would be expected for effects
resulting from the uncertainty principle.

In  terms of  the evanescent  action $u$  in (\ref{eq:u})  and  in the
high-temperature limit $T \gg M$,  we may show quantitatively that the
frequencies of these non-Markovian oscillations are given by
\begin{subequations}
  \begin{align}
    \label{eq:w1}
    &\omega^{(b)}_1(q,u,t)\ =\ q_0\:-\nonumber\\&\frac{\big(q_u^2(t)-
    |\mathbf{q}|^2+m_1^2-m_2^2\big)q_u(t)+b\alpha_{\theta}
    |\mathbf{q}|\lambda^{1/2}\big(q_u^2(t)-
    |\mathbf{q}|^2,m_1^2,m_2^2\big)}{2\big(q_u^2(t)-
    |\mathbf{q}|^2\big)}\;,\\
    &\omega^{(b)}_2(q,u,t)\ =\ \nonumber\\ &\frac{\big(q_u^2(t)-
    |\mathbf{q}|^2-m_1^2+m_2^2\big)q_u(t)-b\alpha_{\theta}
    |\mathbf{q}|\lambda^{1/2}\big(q_u^2(t)-
    |\mathbf{q}|^2,m_1^2,m_2^2\big)}
    {2\big(q_u^2(t)-|\mathbf{q}|^2\big)}\;,
\end{align}
\end{subequations}
with $b,\ \alpha_{\theta}\:=\:\pm 1$,
$\lambda(x,y,z)\:=\:(x^2-y^2-z^2)^2-4y^2z^2$. In addition, we have
introduced the evanescent energy
\begin{equation}
  q_u(t)\ \equiv\ q_0\:-\:\frac{u}{t}
\end{equation}
and we have quoted the result with different masses $m_1$ and $m_2$,
for generality.  Notice that in the limit $t\:\to\:\infty$,
$q_u(t)\:\to\:q_0$ and we obtain the usual time-independent
kinematics.  To the best of our knowledge, such a quantitative
analysis of the non-Markovian evolution of memory effects has not been
reported previously in the literature.

\section{Loopwise-Truncated Time Evolution Equations}

Truncating  the  master  time  evolution  equation  (\ref{eq:evo})  to
leading  order in  a perturbative  loopwise expansion,  we  obtain the
following  one-loop  transport  equation  for the  $\Phi$  statistical
distribution function:
\begin{align}
  \label{eq:phievo}
  &\partial_tf_{\Phi}(|\mathbf{q}|,t)\ =\ 
  -\frac{g^2}{2}\sum_{\alpha,\,\alpha_1,\,\alpha_2}
  \int\!\!\frac{\D{3}{\mathbf{k}}}{(2\pi)^3}\frac{1}{2E_{\Phi}(\mathbf{q})}
  \frac{1}{2E_{\chi}(\mathbf{k})}\frac{1}{2E_{\chi}(\mathbf{q}-\mathbf{k})}
  \nonumber\\&\qquad \times \:
  \frac{t}{2\pi}\,\mathrm{sinc}
  \Big[\Big(\alpha E_{\Phi}(\mathbf{q})-\alpha_1E_{\chi}(\mathbf{k})
    -\alpha_2E_{\chi}(\mathbf{q}-\mathbf{k})\Big)t/2\Big]
  \nonumber\\&\qquad\times\:
  \Big\{\pi+2\mathrm{Si}\Big[\Big(\alpha
    E_{\Phi}(\mathbf{q})+\alpha_1 E_{\chi}(\mathbf{k})+\alpha_2
    E_{\chi}(\mathbf{q}-\mathbf{k})\Big)t/2\Big]\Big\}\nonumber\\&\qquad
  \times\:
  \big\{\big[\theta(-\alpha)+f_{\Phi}(|\mathbf{q}|,t)\big]\nonumber\\&\qquad
  \qquad\times\:\big[\theta(\alpha_1)\big(1+f_{\chi}(|\mathbf{k}|,t)\big)
    +\theta(-\alpha_1)f_{\chi}^C(|\mathbf{k}|,t)\big]\nonumber\\&\quad
  \quad \quad \quad \times \:
  \big[\theta(\alpha_2)\big(1+f^C_{\chi}(|\mathbf{q}-\mathbf{k}|,t)\big)
    +\theta(-\alpha_2)f_{\chi}(|\mathbf{q}-\mathbf{k}|,t)\big]
  \nonumber\\&\qquad -\:
  \big[\theta(\alpha)+f_{\Phi}(|\mathbf{q}|,t)\big]\nonumber\\&\qquad \qquad
  \times\:\big[\theta(\alpha_1)f_{\chi}(|\mathbf{k}|,t)
    +\theta(-\alpha_1)\big(1+f_{\chi}^C(|\mathbf{k}|,t)\big)\big]
  \nonumber\\&\quad\quad \quad \quad \times \:
  \big[\theta(\alpha_2)f^C_{\chi}(|\mathbf{q}-\mathbf{k}|,t)
    +\theta(-\alpha_2)\big(1+f_{\chi}(|\mathbf{q}-\mathbf{k}|,t)\big)\big]
  \big\}\;,
\end{align}
where $\alpha,\ \alpha_1,\ \alpha_2\: = \: \pm 1$. The second and
third lines of (\ref{eq:phievo}) encode the early-time violation of
energy conservation. Replacing these lines by the Markovian
approximation
\begin{equation}
  2\pi\theta(\alpha)\delta\big(E_{\Phi}(\mathbf{q})
  -\alpha_1E_1(\mathbf{k})-\alpha_2E_2(\mathbf{q}-\mathbf{k})\big)\;,
\end{equation}
we recover the semi-classical  Boltzmann equation.  However, given the
equilibrium   initial  conditions  of   our  model,   this  artificial
imposition  of energy conservation  along with  the properties  of the
Bose--Einstein distribution  ensure that the  RHS of (\ref{eq:phievo})
is  zero for all  times. Thus,  the semi-classical  Boltzmann equation
cannot describe  the re-thermalization of  our simple model.   This is
true  also for  gradient expansions  of Kadanoff--Baym  equations when
truncated to zeroth order in time derivatives.  Hence, it is only when
energy-violating effects are systematically considered, as in this new
perturbative approach  with all gradients included,  that the dynamics
of this re-thermalization is properly captured.

It is  clear that (\ref{eq:phievo})  describes only decay  and inverse
decay      processes     in      the      topologies     shown      in
Figure~\ref{fig:kins}.   However,   higher-multiplicity   decays   and
scatterings  can   be  systematically  incorporated   by  consistently
truncating the  master time evolution equation in  (\ref{eq:evo}) to a
higher number of loops.

\section{Conclusions}
\label{sec:conc}

We have  obtained master time evolution equations  for particle number
densities that are  valid to all orders in  perturbation theory and to
all orders  in gradient expansion. The  underlying perturbation series
are  built  from   non-homogeneous  free  propagators  and  explicitly
time-dependent vertices.   Due to  the systematic treatment  of finite
boundary and observation times,  these diagrammatic series remain free
of  pinch singularities  for  all  times.  We  are  therefore able  to
truncate  the  time evolution  equations  in  a perturbative  loopwise
sense, whilst  keeping all orders in gradient  expansion and capturing
the  dynamics on all  timescales. This  includes the  prompt transient
behaviour,  which we have  shown to  be dominated  by energy-violating
processes that  lead to non-Markovian evolution of  memory effects. By
virtue  of  our approach,  we  have been  able  to  provide the  first
quantitative analysis of these memory effects.

The foreseeable  applications of  this new formalism  span high-energy
physics,  astro-particle  physics,   cosmology  and  condensed  matter
physics. Dedicated studies of such applications will be the subject of
future works.

\section*{Acknowledgments}

The  work  of  PM and  AP  is  supported  in  part by  the  \linebreak
Lancaster--Manchester--Sheffield  Consortium  for Fundamental  Physics
under STFC  grant ST/J000418/1.  AP also  acknowledges partial support
by an IPPP associateship from Durham University.

\begin{flushleft}
  \bibliographystyle{model1a-num-names}
  \bibliography{pTFTlett}

\begin{thebibliography}{56}
\expandafter\ifx\csname natexlab\endcsname\relax\def\natexlab#1{#1}\fi
\providecommand{\url}[1]{\texttt{#1}}
\providecommand{\href}[2]{#2}
\providecommand{\path}[1]{#1}
\providecommand{\DOIprefix}{doi:}
\providecommand{\ArXivprefix}{arXiv:}
\providecommand{\URLprefix}{URL: }
\providecommand{\Pubmedprefix}{pmid:}
\providecommand{\doi}[1]{\href{http://dx.doi.org/#1}{\path{#1}}}
\providecommand{\Pubmed}[1]{\href{pmid:#1}{\path{#1}}}
\providecommand{\bibinfo}[2]{#2}
\ifx\xfnm\relax \def\xfnm[#1]{\unskip,\space#1}\fi
\bibitem[{Millington and Pilaftsis(2012)}]{Millington:2012pf}
\bibinfo{author}{P.~Millington}, \bibinfo{author}{A.~Pilaftsis}
  (\bibinfo{year}{2012}). \href{http://arxiv.org/abs/1211.3152\normalfont}{\tt
  arXiv:1211.3152\normalfont}.
\bibitem[{Kolb and Wolfram(1980)}]{Kolb:1979qa}
\bibinfo{author}{E.~W. Kolb}, \bibinfo{author}{S.~Wolfram},
  \bibinfo{journal}{Nucl.Phys.} \bibinfo{volume}{B172} (\bibinfo{year}{1980})
  \bibinfo{pages}{224--284}.
\bibitem[{Carena et~al.(2003)Carena, Quiros, Seco, and Wagner}]{Carena:2002ss}
\bibinfo{author}{M.~S. Carena}, \bibinfo{author}{M.~Quiros},
  \bibinfo{author}{M.~Seco}, \bibinfo{author}{C.~E.~M. Wagner},
  \bibinfo{journal}{Nucl.Phys.} \bibinfo{volume}{B650} (\bibinfo{year}{2003})
  \bibinfo{pages}{24--42}.
\bibitem[{Giudice et~al.(2004)Giudice, Notari, Raidal, Riotto, and
  Strumia}]{Giudice:2003jh}
\bibinfo{author}{G.~F. Giudice}, \bibinfo{author}{A.~Notari},
  \bibinfo{author}{M.~Raidal}, \bibinfo{author}{A.~Riotto},
  \bibinfo{author}{A.~Strumia}, \bibinfo{journal}{Nucl.Phys.}
  \bibinfo{volume}{B685} (\bibinfo{year}{2004}) \bibinfo{pages}{89--149}.
\bibitem[{Pilaftsis and Underwood(2004)}]{Pilaftsis:2003gt}
\bibinfo{author}{A.~Pilaftsis}, \bibinfo{author}{T.~E.~J. Underwood},
  \bibinfo{journal}{Nucl.Phys.} \bibinfo{volume}{B692} (\bibinfo{year}{2004})
  \bibinfo{pages}{303--345}.
\bibitem[{Buchm{\"{u}}ller et~al.(2005)Buchm{\"{u}}ller, Di~Bari, and
  Pl{\"{u}}macher}]{Buchmuller:2004nz}
\bibinfo{author}{W.~Buchm{\"{u}}ller}, \bibinfo{author}{P.~Di~Bari},
  \bibinfo{author}{M.~Pl{\"{u}}macher}, \bibinfo{journal}{Annals Phys.}
  \bibinfo{volume}{315} (\bibinfo{year}{2005}) \bibinfo{pages}{305--351}.
\bibitem[{Pilaftsis and Underwood(2005)}]{Pilaftsis:2005rv}
\bibinfo{author}{A.~Pilaftsis}, \bibinfo{author}{T.~E.~J. Underwood},
  \bibinfo{journal}{Phys.Rev.} \bibinfo{volume}{D72} (\bibinfo{year}{2005})
  \bibinfo{pages}{113001}.
\bibitem[{Davidson et~al.(2008)Davidson, Nardi, and Nir}]{Davidson:2008bu}
\bibinfo{author}{S.~Davidson}, \bibinfo{author}{E.~Nardi},
  \bibinfo{author}{Y.~Nir}, \bibinfo{journal}{Phys.Rep.} \bibinfo{volume}{466}
  (\bibinfo{year}{2008}) \bibinfo{pages}{105--177}.
\bibitem[{Blanchet et~al.(2013)Blanchet, Di~Bari, Jones, and
  Marzola}]{Blanchet:2011xq}
\bibinfo{author}{S.~Blanchet}, \bibinfo{author}{P.~Di~Bari},
  \bibinfo{author}{D.~A. Jones}, \bibinfo{author}{L.~Marzola},
  \bibinfo{journal}{JCAP} \bibinfo{volume}{1301} (\bibinfo{year}{2013})
  \bibinfo{pages}{041}.
\bibitem[{Danielewicz(1984)}]{Danielewicz:1982kk}
\bibinfo{author}{P.~Danielewicz}, \bibinfo{journal}{Annals Phys.}
  \bibinfo{volume}{152} (\bibinfo{year}{1984}) \bibinfo{pages}{239--304}.
\bibitem[{Lipavsk{\'{y}} et~al.(1986)Lipavsk{\'{y}}, {\v{S}}pi{ \v{c}}ka, and
  Velick{\'{y}}}]{Lipavsky:1986zz}
\bibinfo{author}{P.~Lipavsk{\'{y}}}, \bibinfo{author}{V.~{\v{S}}pi{ \v{c}}ka},
  \bibinfo{author}{B.~Velick{\'{y}}}, \bibinfo{journal}{Phys.Rev.}
  \bibinfo{volume}{B34} (\bibinfo{year}{1986}) \bibinfo{pages}{6933--6942}.
\bibitem[{Berera et~al.(1998)Berera, Gleiser, and Ramos}]{Berera:1998gx}
\bibinfo{author}{A.~Berera}, \bibinfo{author}{M.~Gleiser},
  \bibinfo{author}{R.~O. Ramos}, \bibinfo{journal}{Phys.Rev.}
  \bibinfo{volume}{D58} (\bibinfo{year}{1998}) \bibinfo{pages}{123508}.
\bibitem[{Boyanovsky et~al.(1998)Boyanovsky, de~Vega, Holman, Kumar, and
  Pisarski}]{Boyanovsky:1998pg}
\bibinfo{author}{D.~Boyanovsky}, \bibinfo{author}{H.~J. de~Vega},
  \bibinfo{author}{R.~Holman}, \bibinfo{author}{S.~P. Kumar},
  \bibinfo{author}{R.~D. Pisarski}, \bibinfo{journal}{Phys.Rev.}
  \bibinfo{volume}{D58} (\bibinfo{year}{1998}) \bibinfo{pages}{125009}.
\bibitem[{Niegawa(1999)}]{Niegawa:1999pn}
\bibinfo{author}{A.~Niegawa}, \bibinfo{journal}{Prog.Theor.Phys.}
  \bibinfo{volume}{102} (\bibinfo{year}{1999}) \bibinfo{pages}{1--27}.
\bibitem[{Cassing and Juchem(2000)}]{Cassing:1999wx}
\bibinfo{author}{W.~Cassing}, \bibinfo{author}{S.~Juchem},
  \bibinfo{journal}{Nucl.Phys.} \bibinfo{volume}{A665} (\bibinfo{year}{2000})
  \bibinfo{pages}{377--400}.
\bibitem[{Dadi{\'{c}}(2000)}]{Dadic:1999bp}
\bibinfo{author}{I.~Dadi{\'{c}}}, \bibinfo{journal}{Phys.Rev.}
  \bibinfo{volume}{D63} (\bibinfo{year}{2000}) \bibinfo{pages}{025011}.
\bibitem[{Ivanov et~al.(2000)Ivanov, Knoll, and Voskresensky}]{Ivanov:1999tj}
\bibinfo{author}{Y.~Ivanov}, \bibinfo{author}{J.~Knoll},
  \bibinfo{author}{D.~Voskresensky}, \bibinfo{journal}{Nucl.Phys.}
  \bibinfo{volume}{A672} (\bibinfo{year}{2000}) \bibinfo{pages}{313--356}.
\bibitem[{Buchm{\"{u}}ller and Fredenhagen(2000)}]{Buchmuller:2000nd}
\bibinfo{author}{W.~Buchm{\"{u}}ller}, \bibinfo{author}{S.~Fredenhagen},
  \bibinfo{journal}{Phys.Lett.} \bibinfo{volume}{B483} (\bibinfo{year}{2000})
  \bibinfo{pages}{217--224}.
\bibitem[{Morawetz et~al.(2001)Morawetz, Bonitz, Morozov, Ropke, and
  Kremp}]{Morawetz:1999bv}
\bibinfo{author}{K.~Morawetz}, \bibinfo{author}{M.~Bonitz},
  \bibinfo{author}{V.~Morozov}, \bibinfo{author}{G.~Ropke},
  \bibinfo{author}{D.~Kremp}, \bibinfo{journal}{Phys.Rev.}
  \bibinfo{volume}{E63} (\bibinfo{year}{2001}) \bibinfo{pages}{020102}.
\bibitem[{Aarts and Berges(2001)}]{Aarts:2001qa}
\bibinfo{author}{G.~Aarts}, \bibinfo{author}{J.~Berges},
  \bibinfo{journal}{Phys.Rev.} \bibinfo{volume}{D64} (\bibinfo{year}{2001})
  \bibinfo{pages}{105010}.
\bibitem[{Blaizot and Iancu(2002)}]{Blaizot:2001nr}
\bibinfo{author}{J.-P. Blaizot}, \bibinfo{author}{E.~Iancu},
  \bibinfo{journal}{Phys.Rep.} \bibinfo{volume}{359} (\bibinfo{year}{2002})
  \bibinfo{pages}{355--528}.
\bibitem[{Juchem et~al.(2004)Juchem, Cassing, and Greiner}]{Juchem:2003bi}
\bibinfo{author}{S.~Juchem}, \bibinfo{author}{W.~Cassing},
  \bibinfo{author}{C.~Greiner}, \bibinfo{journal}{Phys.Rev.}
  \bibinfo{volume}{D69} (\bibinfo{year}{2004}) \bibinfo{pages}{025006}.
\bibitem[{Prokopec et~al.(2004{\natexlab{a}})Prokopec, Schmidt, and
  Weinstock}]{Prokopec:2003pj}
\bibinfo{author}{T.~Prokopec}, \bibinfo{author}{M.~G. Schmidt},
  \bibinfo{author}{S.~Weinstock}, \bibinfo{journal}{Annals Phys.}
  \bibinfo{volume}{314} (\bibinfo{year}{2004}{\natexlab{a}})
  \bibinfo{pages}{208--265}.
\bibitem[{Prokopec et~al.(2004{\natexlab{b}})Prokopec, Schmidt, and
  Weinstock}]{Prokopec:2004ic}
\bibinfo{author}{T.~Prokopec}, \bibinfo{author}{M.~G. Schmidt},
  \bibinfo{author}{S.~Weinstock}, \bibinfo{journal}{Annals Phys.}
  \bibinfo{volume}{314} (\bibinfo{year}{2004}{\natexlab{b}})
  \bibinfo{pages}{267--320}.
\bibitem[{Berges(2005)}]{Berges:2004yj}
\bibinfo{author}{J.~Berges}, \bibinfo{journal}{AIP Conf.Proc.}
  \bibinfo{volume}{739} (\bibinfo{year}{2005}) \bibinfo{pages}{3--62}.
\bibitem[{Arrizabalaga et~al.(2005)Arrizabalaga, Smit, and
  Tranberg}]{Arrizabalaga:2005tf}
\bibinfo{author}{A.~Arrizabalaga}, \bibinfo{author}{J.~Smit},
  \bibinfo{author}{A.~Tranberg}, \bibinfo{journal}{Phys.Rev.}
  \bibinfo{volume}{D72} (\bibinfo{year}{2005}) \bibinfo{pages}{025014}.
\bibitem[{Berges et~al.(2005)Berges, Bors{\'{a}}nyi, and
  Wetterich}]{Berges:2005ai}
\bibinfo{author}{J.~Berges}, \bibinfo{author}{S.~Bors{\'{a}}nyi},
  \bibinfo{author}{C.~Wetterich}, \bibinfo{journal}{Nucl.Phys.}
  \bibinfo{volume}{B727} (\bibinfo{year}{2005}) \bibinfo{pages}{244--263}.
\bibitem[{Lindner and M{\"{u}}ller(2006)}]{Lindner:2005kv}
\bibinfo{author}{M.~Lindner}, \bibinfo{author}{M.~M. M{\"{u}}ller},
  \bibinfo{journal}{Phys.Rev.} \bibinfo{volume}{D73} (\bibinfo{year}{2006})
  \bibinfo{pages}{125002}.
\bibitem[{Fillion-Gourdeau et~al.(2006)Fillion-Gourdeau, Gagnon, and
  Jeon}]{FillionGourdeau:2006hi}
\bibinfo{author}{F.~Fillion-Gourdeau}, \bibinfo{author}{J.-S. Gagnon},
  \bibinfo{author}{S.~Jeon}, \bibinfo{journal}{Phys.Rev.} \bibinfo{volume}{D74}
  (\bibinfo{year}{2006}) \bibinfo{pages}{025010}.
\bibitem[{De~Simone and Riotto(2007)}]{DeSimone:2007rw}
\bibinfo{author}{A.~De~Simone}, \bibinfo{author}{A.~Riotto},
  \bibinfo{journal}{JCAP} \bibinfo{volume}{0708} (\bibinfo{year}{2007})
  \bibinfo{pages}{002}.
\bibitem[{Cirigliano et~al.(2008)Cirigliano, De~Simone, Isidori, Masina, and
  Riotto}]{Cirigliano:2007hb}
\bibinfo{author}{V.~Cirigliano}, \bibinfo{author}{A.~De~Simone},
  \bibinfo{author}{G.~Isidori}, \bibinfo{author}{I.~Masina},
  \bibinfo{author}{A.~Riotto}, \bibinfo{journal}{JCAP} \bibinfo{volume}{0801}
  (\bibinfo{year}{2008}) \bibinfo{pages}{004}.
\bibitem[{Garbrecht and Konstandin(2009)}]{Garbrecht:2008cb}
\bibinfo{author}{B.~Garbrecht}, \bibinfo{author}{T.~Konstandin},
  \bibinfo{journal}{Phys.Rev.} \bibinfo{volume}{D79} (\bibinfo{year}{2009})
  \bibinfo{pages}{085003}.
\bibitem[{Garny et~al.(2010)Garny, Hohenegger, Kartavtsev, and
  Lindner}]{Garny:2009qn}
\bibinfo{author}{M.~Garny}, \bibinfo{author}{A.~Hohenegger},
  \bibinfo{author}{A.~Kartavtsev}, \bibinfo{author}{M.~Lindner},
  \bibinfo{journal}{Phys.Rev.} \bibinfo{volume}{D81} (\bibinfo{year}{2010})
  \bibinfo{pages}{085027}.
\bibitem[{Cirigliano et~al.(2010)Cirigliano, Lee, Ramsey-Musolf, and
  Tulin}]{Cirigliano:2009yt}
\bibinfo{author}{V.~Cirigliano}, \bibinfo{author}{C.~Lee},
  \bibinfo{author}{M.~J. Ramsey-Musolf}, \bibinfo{author}{S.~Tulin},
  \bibinfo{journal}{Phys.Rev.} \bibinfo{volume}{D81} (\bibinfo{year}{2010})
  \bibinfo{pages}{103503}.
\bibitem[{Beneke et~al.(2011)Beneke, Garbrecht, Fidler, Herranen, and
  Schwaller}]{Beneke:2010dz}
\bibinfo{author}{M.~Beneke}, \bibinfo{author}{B.~Garbrecht},
  \bibinfo{author}{C.~Fidler}, \bibinfo{author}{M.~Herranen},
  \bibinfo{author}{P.~Schwaller}, \bibinfo{journal}{Nucl.Phys.}
  \bibinfo{volume}{B843} (\bibinfo{year}{2011}) \bibinfo{pages}{177--212}.
\bibitem[{Anisimov et~al.(2011)Anisimov, Buchm{\"{u}}ller, Drewes, and
  Mendizabal}]{Anisimov:2010dk}
\bibinfo{author}{A.~Anisimov}, \bibinfo{author}{W.~Buchm{\"{u}}ller},
  \bibinfo{author}{M.~Drewes}, \bibinfo{author}{S.~Mendizabal},
  \bibinfo{journal}{Annals Phys.} \bibinfo{volume}{326} (\bibinfo{year}{2011})
  \bibinfo{pages}{1998--2038}.
\bibitem[{Hamaguchi et~al.(2012)Hamaguchi, Moroi, and
  Mukaida}]{Hamaguchi:2011jy}
\bibinfo{author}{K.~Hamaguchi}, \bibinfo{author}{T.~Moroi},
  \bibinfo{author}{K.~Mukaida}, \bibinfo{journal}{JHEP} \bibinfo{volume}{1201}
  (\bibinfo{year}{2012}) \bibinfo{pages}{083}.
\bibitem[{Fidler et~al.(2012)Fidler, Herranen, Kainulainen, and
  Rahkila}]{Fidler:2011yq}
\bibinfo{author}{C.~Fidler}, \bibinfo{author}{M.~Herranen},
  \bibinfo{author}{K.~Kainulainen}, \bibinfo{author}{P.~M. Rahkila},
  \bibinfo{journal}{JHEP} \bibinfo{volume}{1202} (\bibinfo{year}{2012})
  \bibinfo{pages}{065}.
\bibitem[{Gautier and Serreau(2012)}]{Gautier:2012vh}
\bibinfo{author}{F.~Gautier}, \bibinfo{author}{J.~Serreau},
  \bibinfo{journal}{Phys.Rev.} \bibinfo{volume}{D86} (\bibinfo{year}{2012})
  \bibinfo{pages}{125002}.
\bibitem[{Drewes et~al.(2013)Drewes, Mendizabal, and Weniger}]{Drewes:2012qw}
\bibinfo{author}{M.~Drewes}, \bibinfo{author}{S.~Mendizabal},
  \bibinfo{author}{C.~Weniger}, \bibinfo{journal}{Phys.Lett.}
  \bibinfo{volume}{B718} (\bibinfo{year}{2013}) \bibinfo{pages}{1119--1124}.
\bibitem[{Kadanoff and Baym(1989)}]{Kadanoff1989}
\bibinfo{author}{L.~P. Kadanoff}, \bibinfo{author}{G.~Baym},
  \bibinfo{title}{Quantum statistical mechanics},
  \bibinfo{publisher}{Addison-Wesley}, \bibinfo{year}{1989}.
\bibitem[{Gasenzer et~al.(2010)Gasenzer, Kessler, and
  Pawlowski}]{Gasenzer:2010rq}
\bibinfo{author}{T.~Gasenzer}, \bibinfo{author}{S.~Kessler},
  \bibinfo{author}{J.~M. Pawlowski}, \bibinfo{journal}{Eur.Phys.J.}
  \bibinfo{volume}{C70} (\bibinfo{year}{2010}) \bibinfo{pages}{423--443}.
\bibitem[{Sigl and Raffelt(1993)}]{Sigl:1992fn}
\bibinfo{author}{G.~Sigl}, \bibinfo{author}{G.~Raffelt},
  \bibinfo{journal}{Nucl.Phys.} \bibinfo{volume}{B406} (\bibinfo{year}{1993})
  \bibinfo{pages}{423--451}.
\bibitem[{Gagnon and Shaposhnikov(2011)}]{Gagnon:2010kt}
\bibinfo{author}{J.-S. Gagnon}, \bibinfo{author}{M.~Shaposhnikov},
  \bibinfo{journal}{Phys.Rev.} \bibinfo{volume}{D83} (\bibinfo{year}{2011})
  \bibinfo{pages}{065021}.
\bibitem[{Winter(1985)}]{Winter:1986da}
\bibinfo{author}{J.~Winter}, \bibinfo{journal}{Phys.Rev.} \bibinfo{volume}{D32}
  (\bibinfo{year}{1985}) \bibinfo{pages}{1871--1888}.
\bibitem[{Bornath et~al.(1996)Bornath, Kremp, Kraeft, and
  Schlanges}]{Bornath:1996zz}
\bibinfo{author}{T.~Bornath}, \bibinfo{author}{D.~Kremp},
  \bibinfo{author}{W.~D. Kraeft}, \bibinfo{author}{M.~Schlanges},
  \bibinfo{journal}{Phys.Rev.} \bibinfo{volume}{E54} (\bibinfo{year}{1996})
  \bibinfo{pages}{3274--3284}.
\bibitem[{Schwinger(1961)}]{Schwinger:1960qe}
\bibinfo{author}{J.~S. Schwinger}, \bibinfo{journal}{J.Math.Phys.}
  \bibinfo{volume}{2} (\bibinfo{year}{1961}) \bibinfo{pages}{407--432}.
\bibitem[{Keldysh(1964)}]{Keldysh:1964ud}
\bibinfo{author}{L.~Keldysh}, \bibinfo{journal}{Zh.Eksp.Teor.Fiz.}
  \bibinfo{volume}{47} (\bibinfo{year}{1964}) \bibinfo{pages}{1515--1527}.
\bibitem[{Calzetta and Hu(1987)}]{Calzetta:1986ey}
\bibinfo{author}{E.~Calzetta}, \bibinfo{author}{B.~L. Hu},
  \bibinfo{journal}{Phys.Rev.} \bibinfo{volume}{D35} (\bibinfo{year}{1987})
  \bibinfo{pages}{495--509}.
\bibitem[{Calzetta and Hu(1988)}]{Calzetta:1986cq}
\bibinfo{author}{E.~Calzetta}, \bibinfo{author}{B.~L. Hu},
  \bibinfo{journal}{Phys.Rev.} \bibinfo{volume}{D37} (\bibinfo{year}{1988})
  \bibinfo{pages}{2878--2900}.
\bibitem[{Cornwall et~al.(1974)Cornwall, Jackiw, and
  Tomboulis}]{Cornwall:1974vz}
\bibinfo{author}{J.~M. Cornwall}, \bibinfo{author}{R.~Jackiw},
  \bibinfo{author}{E.~Tomboulis}, \bibinfo{journal}{Phys.Rev.}
  \bibinfo{volume}{D10} (\bibinfo{year}{1974}) \bibinfo{pages}{2428--2445}.
\bibitem[{Kolb and Turner(1994)}]{KolbTurner}
\bibinfo{author}{E.~Kolb}, \bibinfo{author}{M.~Turner}, \bibinfo{title}{The
  Early Universe}, \bibinfo{publisher}{Westview Press}, \bibinfo{year}{1994},
  p. \bibinfo{pages}{116}.
\bibitem[{Altherr and Seibert(1994)}]{Altherr:1994fx}
\bibinfo{author}{T.~Altherr}, \bibinfo{author}{D.~Seibert},
  \bibinfo{journal}{Phys.Lett.} \bibinfo{volume}{B333} (\bibinfo{year}{1994})
  \bibinfo{pages}{149--152}.
\bibitem[{Altherr(1995)}]{Altherr:1994jc}
\bibinfo{author}{T.~Altherr}, \bibinfo{journal}{Phys.Lett.}
  \bibinfo{volume}{B341} (\bibinfo{year}{1995}) \bibinfo{pages}{325--331}.
\bibitem[{Greiner and Leupold(1999)}]{Greiner:1998ri}
\bibinfo{author}{C.~Greiner}, \bibinfo{author}{S.~Leupold},
  \bibinfo{journal}{Eur.Phys.J.} \bibinfo{volume}{C8} (\bibinfo{year}{1999})
  \bibinfo{pages}{517--522}.
\bibitem[{Le~Bellac(2000)}]{LeBellac2000}
\bibinfo{author}{M.~Le~Bellac}, \bibinfo{title}{Thermal field theory},
  \bibinfo{publisher}{Cambridge University Press}, \bibinfo{year}{2000},
  p.~\bibinfo{pages}{24}.

\end{thebibliography}
\end{flushleft}

\end{document}